\documentclass[english,aps,pre,showpacs,reprint,nobibnotes]{revtex4-1}

\usepackage{amsmath,amssymb}

\usepackage[utf8x]{inputenc}

\usepackage[T1]{fontenc}
\usepackage{float}
\usepackage{mathtools}
\usepackage{amsmath}
\usepackage{amssymb}
\usepackage{graphicx}
\usepackage{nicefrac}

\begin{document}
	
	\title{A Nonhyperbolic Toy Model of Cochlear Dynamics}
	
	\date{\today}
	
	\author{Keith Hayton}
	\email{khayt86@gmail.com }
	\thanks{author contributed equally}
	\author{Dimitrios Moirogiannis}
	\email{dmoirogi@gmail.com}
	\thanks{author contributed equally}
	\author{Marcelo Magnasco}
	\affiliation{ Center for Study in Physics and Biology, The Rockefeller University}

\begin{abstract}
	Cochlea displays complex and highly nonlinear behavior in response to wide-ranging auditory stimuli. While there have been many recent advancements in the modeling of cochlear dynamics, it remains unclear what mathematical structures underlie the essential features of the extended cochlea. We construct a dynamical system consisting of a series of strongly coupled critical oscillators to show that high-dimensional nonhyperbolic dynamics can account for high-order compressive nonlinearities, amplification of weak input, frequency selectivity, and traveling waves of activity. As a single Hopf bifurcation generically gives rise to features of cochlea at a local level, the nonhyperbolicity mechanism proposed in this paper can be seen as a higher-dimensional analogue for the entire extended cochlea. 
	\end{abstract}

	\maketitle

	\section{Introduction}
	
\begin{figure}[bth]
		\begin{centering}
			\includegraphics[scale=0.4]{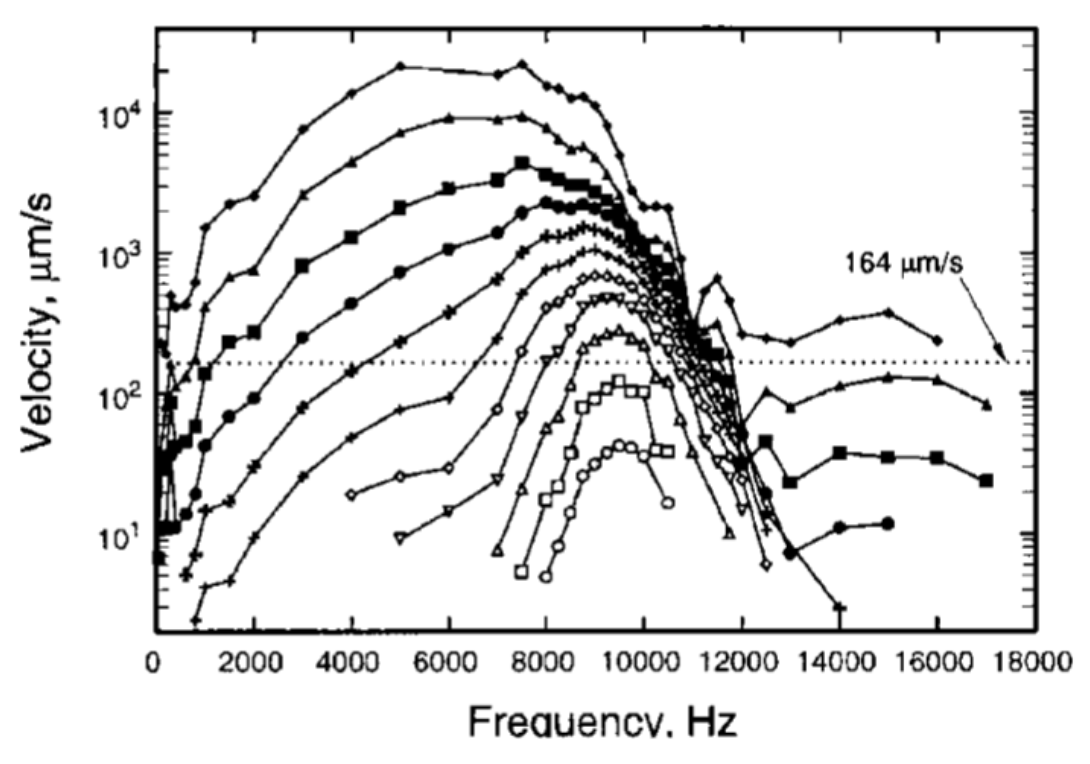}
			\par\end{centering}
		
		\begin{centering}
			\caption{\label{fig:Fig1}\textbf{Cochlear Velocimetric Data}. \cite{Ruggero2000} Cochlear velocimetry data taken by laser interferometry at a spot in the basilar membrane of a living chinchilla. Basilar membrane speed vs frequency for varying intensities is plotted. Adjacent curves are separated by 10 dB.}
			\par\end{centering}
		
	\end{figure}

	The nature of signal processing in the cochlea has been a focus of interest in the study of sound perception. Gold was first to posit that the cochlea does not operate as a passive Fourier transformer but instead utilises an active regenerative system which amplifies incoming signals \cite{Gold1948H1,Gold1948H2}. Experiments on living specimens have confirmed Gold's hypothesis \cite{Ruggero1992,Ruggero2000} despite initial studies on cadavers showing that cochlea acts as a simple passive spatial frequency analyser \cite{Bekesy1960}. Indeed, live, reasonably intact cochlea exhibits an active nonlinear process with three key characteristic properties: high gain amplification, sharp frequency tuning, and nonlinear compression of the dynamic range (Fig.\ \ref{fig:Fig1}) \cite{Ruggero1992,Ruggero1997}.
	
	To understand the origin of the active process, one first needs to examine the physiological and anatomical properties of the cochlea. As sound waves enter the inner ear, they set into vibration the cochlear partition, which results in traveling waves of activity that propagate unidirectionally from base to apex. The \textit{basilar membrane} is the main compliance of the cochlear partition, whose stiffness increases, by two orders of magnitude, from apex to base. This stiffness gradient results in each point on the basilar membrane responding maximally to a single characteristic frequency with high frequencies at the base and exponentially decreasing frequencies towards the apex \cite{Bekesy1960}. 
	The cochlear partition also contains the organ of Corti, where the \textit{hair cells}, the sound sensitive cells, are located. The hair cells play an essential role in generating the features of the active process through \textit{hair-bundle motility} in nonmammals \cite{Hudpseth2008,Hudspeth2000,Martin2008,Martin1999,Martin2001} a combination of hair-bundle motility \cite{Chan2005Dec,Chan2005Feb,Kennedy2005,Kennedy2006}  and \textit{membrane-based electromotility} in mammals \cite{Ashmore2008,Dallos2006,Fettiplace2006}. 
	
	It is now known \cite{Hudspeth2010} that at a local level, at a specific location on the basilar membrane, the characteristic features of the active process generically arise from a \textit{Hopf bifurcation} \cite{Wiggins2003introduction, Kuznetsov2013}, a form of structural instability in dynamical systems theory which gives rise to critical oscillations. Several theoretical studies have extensively investigated this relationship between local cochlear dynamics and a single Hopf bifurcation \cite{Choe1998,Eguiluz2000,Camalet2000}. 
	
	At the level of the extended cochlea, encompassing the full basilar membrane, models relying on multiple Hopf bifurcations, biophysical details of basilar membrane mechanics, and surrounding fluid hydrodyanmics have successfully reproduced a number of key features of cochlear dynamics \cite{Kern2003,Duke2003,Magnasco2003}. However, in contrast to a single Hopf bifurcation in the local case, it is still unclear what underlying high-dimensional mathematical structures can generically give rise to the complex, nonlinear responses of the extended cochlea. In this paper, we propose such a mathematical structure; we show that a high-dimensional nonhyperbolic system residing on a full-dimensional center manifold can exhibit the key nonlinear features of the active process as described above. 
	
	In dynamical systems theory, the classical approach to studying behavior in the neighborhood of an equilibrium point is to examine the eigenvalues of the Jacobian of the system at this point. If all eigenvalues have nonzero real part, then the equilibrium point is called \textit{hyperbolic} and the dynamics around the point is topologically conjugate to the linearized system determined by the Jacobian \cite{Grobman1959,Hartman1960a,Hartman1960b}. On the other hand, if there exists at least one eigenvalue with zero real part, then the equilibrium point is called \textit{nonhyperbolic} and the linearization does not determine the qualitative dynamics around the point. 
	
	Dynamics around nonhyperbolic fixed points are complex and give rise to a number of interesting features. First, since the dynamics are not enslaved by the exponent of the Jacobian, nonlinearities and input parameters play a crucial role in determining dynamical properties such as relaxation timescales and correlations \cite{Yan2012, Hayton2018}. Nonhyperbolic points are also not \textit{structurally stable}, meaning that small perturbations of the vector field can lead to substantial topological changes of the orbits around the point. Examples of these topological changes include the appearance of new invariant sets such as periodic orbits and tori, and if the dimension is high enough, chaotic dimensions can arise.
	
	The standard technique in studying dynamics around a nonhyperbolic point is to investigate the reduced dynamics on an invariant subspace called a \textit{center manifold}. The local existence of center manifolds has been rigorously proven \cite{Kelley1967,Carr2012}, and its dimension is equal to the number of eigenvalues on the imaginary axis. If there are no eigenvalues with positive real part (unstable modes), then the center manifold is attracting \cite{Carr2012}, and instead of studying the full system, we can study the reduced dynamics on the center manifold. The \textit{approximation theorem} for center manifolds provides us with the tool to calculate the center manifold and the reduced dynamics up to any degree of accuracy \cite{Carr2012}.
	
	In this paper, we present a dynamical system poised at a nonhyperbolic point with all eigenvalues of the linearization being purely imaginary; the dynamics reside on a full dimensional center manifold. In general, the greater the number of eigenvalues on the imaginary axis, the more complex the dynamics could be \cite{Wiggins2003introduction}. Thus, the rich and complex behavior we will be discussing in our paper is not surprising. We posit that the nervous system utilizes nonhyperbolic equilibrium points and the corresponding unique dynamical properties on center manifolds to flexibly respond to a wide range of input parameters and exhibit complex nonlinear behavior. Our aim in this paper is not to provide a detailed anatomical and biophysical model of cochlea, but rather to construct a \textit{toy model} which provides an existence proof that center manifold dynamics can account for and connect key characteristic properties of cochlea: high-order compressive nonlinearities, amplification of weak input, frequency selectivity and traveling waves of activity. We do not suggest that a simple 1-D line topology is necessarily present in cochlear anatomy; although, if the cochlea does indeed utilize center manifolds in the processing of sound, it might be the case that the full high dimensional phase space of cochlear dynamics could be reduced to simple, low dimensional structures on the center manifold. This approach of constructing a toy model to explain how a given mechanism can lead to a particular set of properties is common practice in theoretical physics and is the underlying philosophical approach of several well known theoretical neuroscience models, e.g. Wilson-Cowan  equations, Hopfield networks, and Kuramoto models \cite{ermentrout2010mathematical,hoppensteadt2012weakly}. 
	
	There are a number of studies regarding nonhyperbolic dynamics in neural systems, including entire hemisphere ECoG recordings \cite{solovey2015}, experimental studies in premotor and motor cortex \cite{churchland2012}, theoretical \cite{seung1998} and experimental studies \cite{seung2000} of \textit{slow manifolds} (a specific case of center manifolds) in oculomotor control, slow manifolds in decision making \cite{machens2005}, Hopf bifurcation in the olfactory system \cite{freeman2005}, and theoretical work on regulated criticality \cite{bienenstock1998}.
	
	Nonhyperbolic dynamics, also commonly referred to as \textit{dynamical criticality}, is distinct from \textit{statistical criticality} \cite{beggs2012being}, which is related to the statistical mechanics of second-order phase transitions. It has been proposed that neural systems \cite{chialvo2010emergent}, and more generally biological systems \cite{mora2011biological}, are statistically critical in the sense that they are poised near the critical point of a phase transitions \cite{da1998criticality,fraiman2009ising}. Statistical criticality is characterized by power law behavior such as avalanches \cite{beggs2003neuronal,levina2007dynamical,gireesh2008neuronal} and long-range spatiotemporal correlations \cite{eguiluz2005scale,kitzbichler2009broadband}. While both dynamical criticality and statistical criticality have had success in neuroscience, their relation is still far from clear \cite{magnasco2009self,mora2011biological,kanders2017avalanche}.
	
	\section{Methods}
	
	\subsection{Mechanism for Higher-Order Compression}
	
\begin{figure}[bth]
		\begin{centering}
			\includegraphics[scale=0.3]{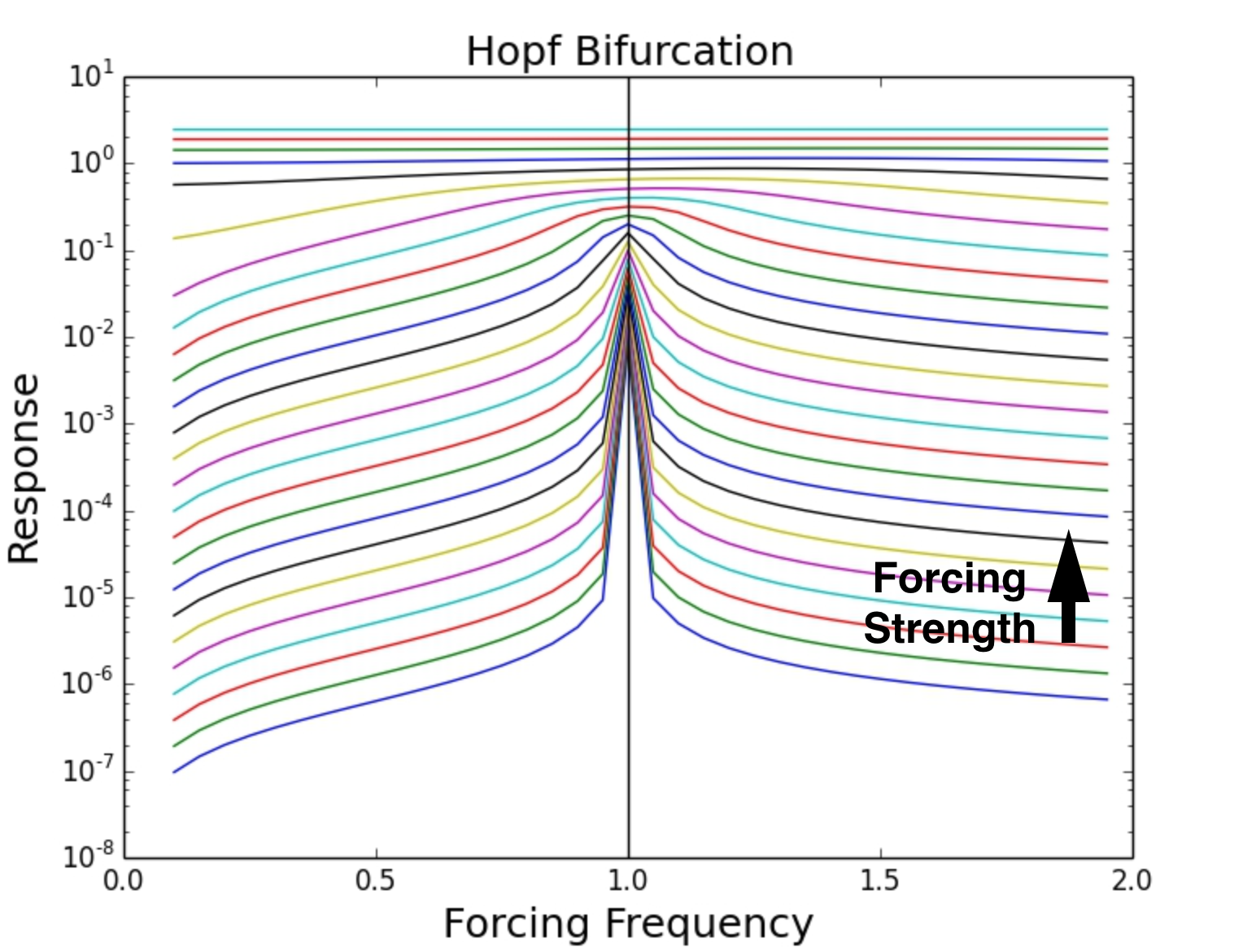}
			\par\end{centering}
		
		\begin{centering}
			\caption{\label{fig:Fig2}\textbf{Response of a Single Critical Oscillator}. The response of a single critical oscillator to a range of forcing frequencies and forcing strengths is shown. Each colored curve corresponds to a different forcing strength and the strength increases from bottom to top (direction of arrow). At resonance, when the forcing frequency is 1.0, the response increases as the cubic root of the forcing strength. Away from resonance the response is linear for small forcing and a cubic root for larger forcing.}
			\par\end{centering}
	\end{figure}

	We first show that nonlinear compression arises naturally from a system of strongly coupled critical oscillators, on an order exponentially larger than one would obtain from a single critical oscillator alone. It's well known that the response of a single Hopf oscillator to exponentially distributed periodic input is given by the curves in Fig.\ \ref{fig:Fig2}. At the center of resonance, the response $R$ scales as the cubic root of the forcing strength $F$, $R\propto{F^{\nicefrac{1}{3}}}$. Away from resonance, the response scales linearly for small forcing and as a cubic root for large forcing. 
	
\begin{figure}[bth]
		\begin{centering}
			\includegraphics[scale=0.45]{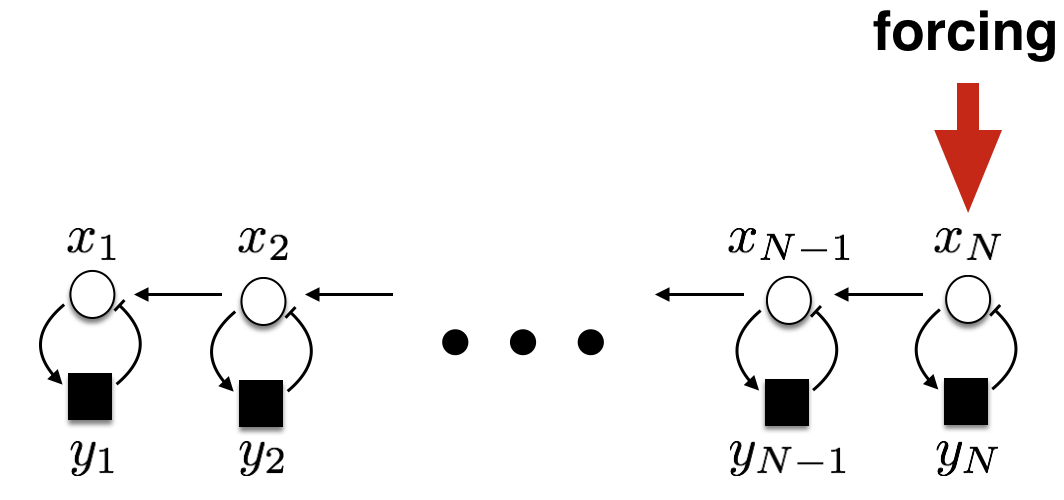}
			\par\end{centering}

		\begin{centering}
			\caption{\label{fig:Fig3}\textbf{Strongly Coupled Critical Oscillators}. Network of $N$ critical oscillators with excitatory cells on top and inhibitory cells on the bottom.  The oscillators are coupled together with connections along the excitatory layer. All connections have strength of magnitude equal to 1. Input to the network is denoted by the arrow.}
			\par\end{centering}
		
	\end{figure}

\begin{figure*}[bth]
		\begin{centering}
			\includegraphics[scale=0.5]{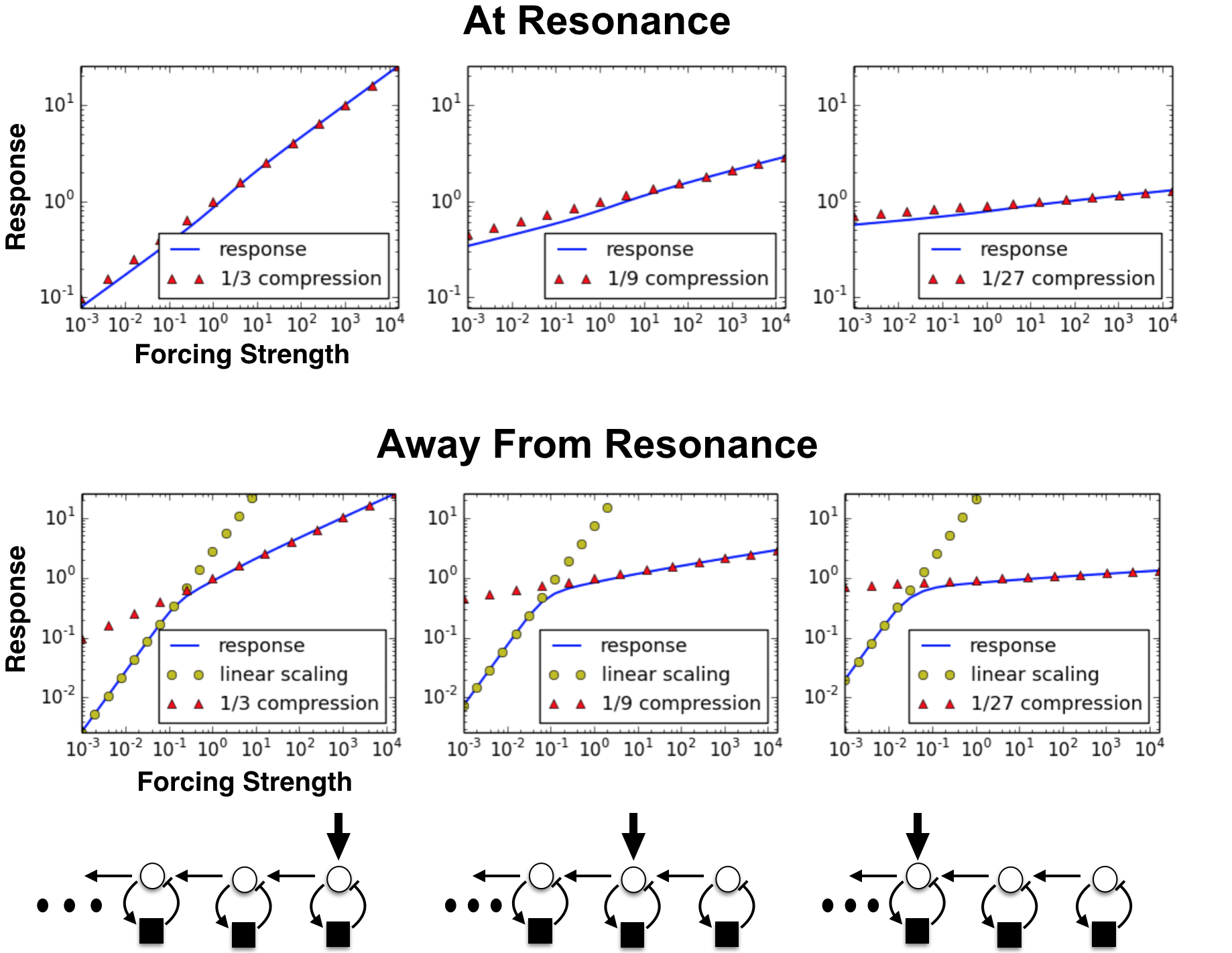}
			\par\end{centering}

		\begin{centering}
			\caption{\label{fig:Fig4}\textbf{Scaling Behavior Along a Series of Coupled Identical Critical Oscillators}. We force the last oscillator in a series of coupled identical critical oscillators with periodic input and plot the response as a function of forcing strength at the three different network locations (vertical arrow).  The top and bottom rows in the figure correspond to the forcing frequency exactly at and away from resonance, respectively. At resonance, the response $|x_{N-d}|$ at a distance $d$ from the input scales as $|x_{N-d}|\propto{F^{\nicefrac{1}{3^{d+1}}}}$ for the entire range of forcing strengths. Away from resonance, the response scales linearly for small forcing and as $|x_{N-d}|\propto{F^{\nicefrac{1}{3^{d+1}}}}$ for large forcing.}
			\par\end{centering}
	\end{figure*}

	Now we look at the more interesting case of a series of unidirectionally coupled Hopf oscillators (Fig.\ \ref{fig:Fig3}); the output of one oscillator acts as input to the next oscillator downstream. For sake of exposition, let us consider the case where all connections have strength of magnitude 1, coupling nonlinearities are discarded, and we force, with periodic input, only the last cell on the top row:
	
	\begin{eqnarray}
	\dot{x}_{i}&=&-y_{i}-|x_{i}|^{2}x_{i}+x_{i+1}+F\delta_{i,N}e^{i{\omega}t} \label{eq:hopfComp} \\
	\dot{y}_{i}&=&x_{i}-|y_{i}|^2y_{i}\nonumber
	\end{eqnarray}

	\noindent
	where $i\in\{1,2,...,N\}$, $x_{i}$ and $y_{i}$ $\in\mathbb{C}$. At resonance, when $\omega=1$, the response $|x_{N-d}|$ at a distance $d$ from the input scales as $|x_{N-d}|\propto{F^{\nicefrac{1}{3^{d+1}}}}$ for all forcing strengths $F$. Away from resonance, the response scales linearly for small forcing and as $|x_{N-d}|\propto{F^{\nicefrac{1}{3^{d+1}}}}$ for large forcing. This scaling behavior is illustrated in Fig.\ \ref{fig:Fig4} where we plot the response of the last cell we force as well as the next two oscillators downstream.
	
	By construction, the linear connectivity matrix describing the network in Fig.\ \ref{fig:Fig3} is non-normal and has purely imaginary eigenvalues. Both features are not necessary to produce the high order power law scaling seen in Fig.\ \ref{fig:Fig4}; however, at resonance, purely imaginary eigenvalues are necessary to give rise to nonlinear compression across all forcing strengths. 
	
	\subsection{Main Toy Model}
\begin{figure*}[bth]
		\begin{centering}
			\includegraphics[scale=0.25]{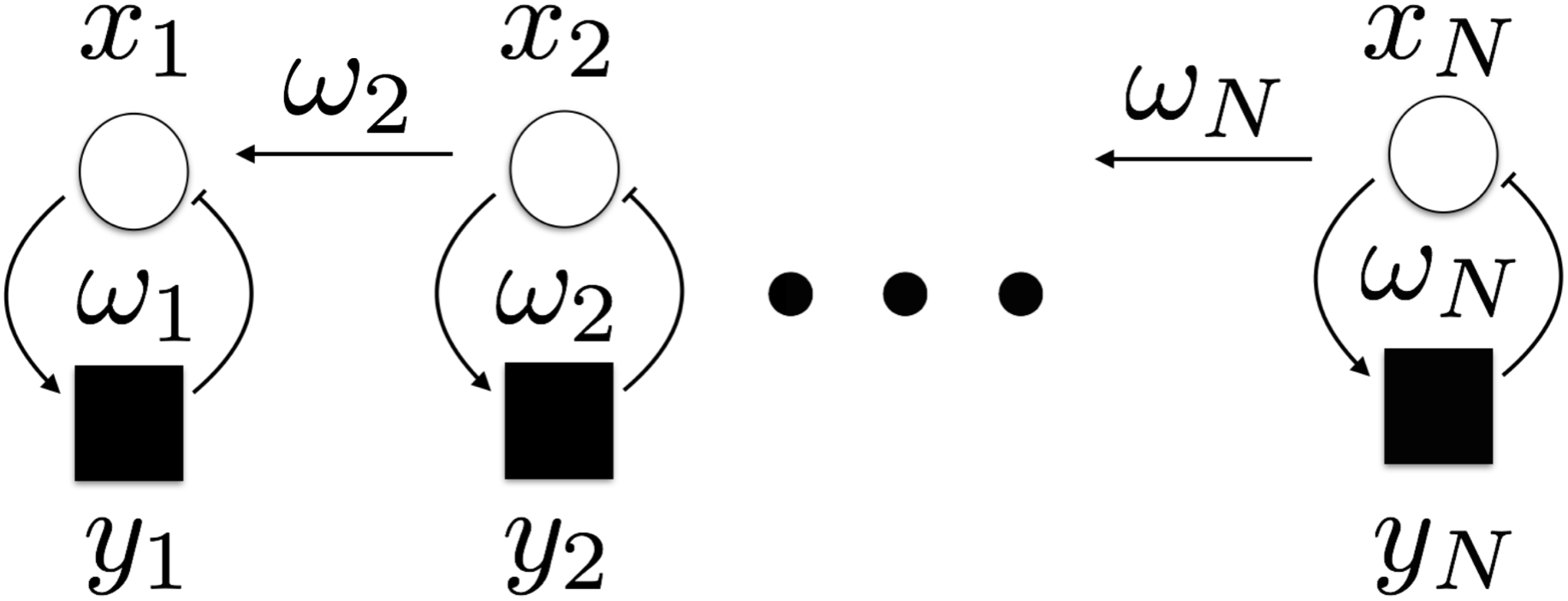}
			\par\end{centering}
		
		\begin{centering}
			\caption{\label{fig:Fig5}\textbf{Network of Coupled Oscillators}. A network of $2N$ cells, with top and bottom cell activities labeled as $x_{i}$ and $y_{i}$, respectively, form $N$ oscillators with exponentially distributed characteristic frequencies $\omega_{1}<\omega_{2}<\omega_{3}<...<\omega_{N}$. The oscillators are coupled to one another through unidirectional connections along the top.}
			\par\end{centering}

	\end{figure*}

	We have shown that a series of coupled critical oscillators can generate responses which are compressed to an arbitrarily large degree. We will now incorporate this mechanism into an extended model of cochlea that is selective to frequencies over an exponential range. Let us consider a network of $2N$ cells paired together to form $N$ strongly coupled oscillators as shown in Fig.\ \ref{fig:Fig5}. The activity of the cells on the top row are given by $x_{i}\in\mathbb{C}$ and cells on the bottom by $y_{i}\in\mathbb{C}$. The $ith$ oscillator in the network, consisting of cells $x_{i}$ and $y_{i}$, has characteristic frequency $\omega_{i}$ and is coupled to the oscillator on its left via a unidirectional connection of strength $\omega_{i}$ from $x_{i}$ to $x_{i-1}$. In agreement with frequency selectivity in the basilar membrane, the characteristic frequencies are exponentially distributed with  $\omega_{1}<\omega_{2}<\omega_{3}<...<\omega_{N}$.
	
	The cells evolve in time, under the influence of forcings $I_{x_{i}}(t)$ and $I_{y_{i}}(t)$, according to the equations:
	
	\begin{eqnarray}
	\dot{x}_{i}&=&-\omega_{i}y_{i}-|x_{i}|^{2}x_{i}+\omega_{i+1}x_{i+1}+I_{x_{i}}(t) \label{eq:hopf} \\
	\dot{y}_{i}&=&\omega_{i}x_{i}-|y_{i}|^2y_{i}+I_{y_{i}}(t)\nonumber
	\end{eqnarray}
	
	\noindent
	where $i\in\{1,2,3,...,N\}$, $x_{i}$ and $y_{i}$ $\in\mathbb{C}$ (for formality let us define: $x_{N+1}=\omega_{N+1}=0$). It should be noted that in these equations, nonlinearities are confined to the cubic order terms and do not couple distinct cells in the network. 
	
	Now let us concatenate the vectors $\vec{x}$ and $\vec{y}$ into a single vector $X\in\mathbb{C}^{2N}$ such that $X_{i}=x_{i}$ and $X_{i+N}=y_{i}$ for $i=1,...,N$. In a similar way, we form $I(t)\in\mathbb{C}^{2N}$ from $I_{x_{i}}(t)$ and $I_{y_{i}}(t)$. We can then rewrite (\ref{eq:hopf}) as: 
	
	\begin{eqnarray}
	\dot{X}&=&AX-|X|^{2}X+I(t)    \label{eq:hopfCompact}
	\end{eqnarray}
	
	\noindent
	where the $|X|^{2}$ and $|X|^{2}X$ are element wise, and the matrix $A$ describes the connectivity in Fig.\ \ref{fig:Fig5}. It is easy to check that $A$ has a purely imaginary spectrum. Thus, the center manifold is full dimensional and we should expect to see nonlinear behavior even for the entire range of forcings.
	
	We are specifically interested in the time-asymptotic response of
	the system to periodic input $I(t)=Fe^{i{\omega}t}$. The high dimensionality and nonhyperbolicity of the dynamics makes the ODE in (\ref{eq:hopfCompact}) difficult to integrate since the step size necessary to numerically integrate the system decreases as the forcing strength is increased. Fortunately, we can bypass numerical integration methods by looking for asymptotic solutions $X(t)=Ze^{i\omega t}$, where $Z\in\mathbb{C}^{2N}$. Substituting these into (\ref{eq:hopfCompact}), we find that:  
	
	\begin{equation}
	0=(A-i\omega)Z-|Z|^{2}Z+F  \label{eq:hopfReduced}
	\end{equation}
	and define $g(Z)$ to be equal to the right hand side of (\ref{eq:hopfReduced}).
	
	The solution of (\ref{eq:hopfReduced}) can be found numerically by using the multivariable
	Newton-Raphson method in $\mathbb{C}^{4N}$:
	
	\begin{equation}
	\widetilde{Z}\rightarrow\widetilde{Z}-J(\widetilde{Z})^{-1}\widetilde{g}(Z)\label{CritFixedPoint}
	\end{equation}
	where $\widetilde{Z}\coloneqq(z_1,\:z_2)=(Re(Z),\:Im(Z))$, $\widetilde{g}(Z)\coloneqq(g_1(Z),\:g_2(Z))=(Re(g(Z)),\:Im(g(Z)))$,
	and $J$ is the Jacobian of $\widetilde{g}$ with respect to $\widetilde{Z}$: $$J_{ij}(z)=\frac{{\partial}g_i}{{\partial}z_j}$$
	
	This Newton-Raphson algorithm can fail to converge for randomly chosen initial points.
	Furthermore, the algorithm's trajectory can get trapped in periodic orbits for step sizes not adequately small.  To make simulations possible, we incorporate adaptive step sizes and initial points into the algorithm.
	
	We define the response of the system as $|Z|$, the element-wise complex modulus of $Z$. We define the response of a single oscillator to be the $l^2$-norm of the oscillator $R_{i}=\sqrt{|Z|_{i}^{2}+|Z|_{i+N}^2}$ for $i\in{1,...,N}$, which makes sense since the response of top and bottom cells are proportional element wise.

	\section{Results}
	
	\subsection{Frequency Tuning, Compression, and Amplification}
\begin{figure*}[bth]
	 	\begin{centering}
	 		\includegraphics[scale=0.5]{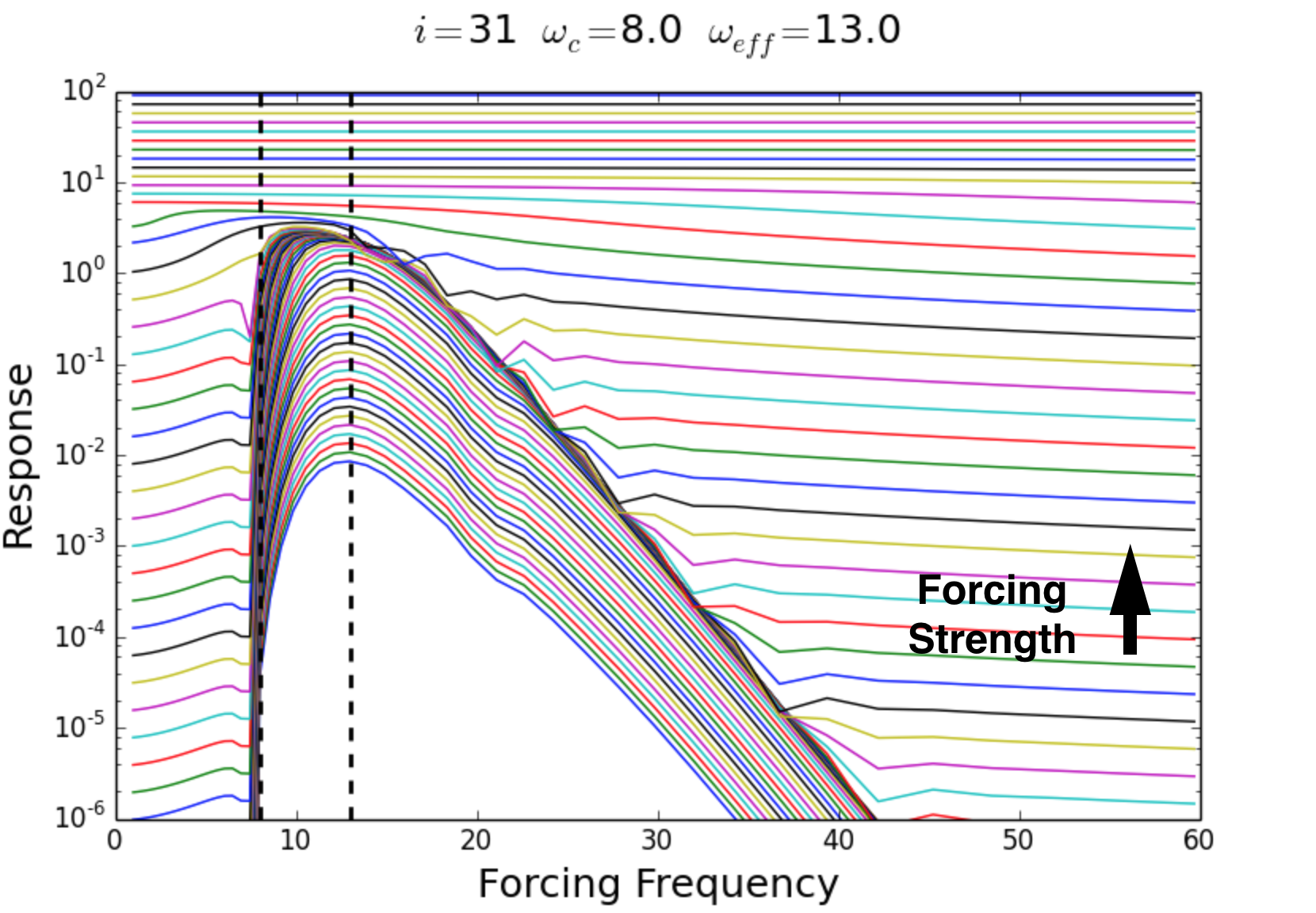}
	 		\par\end{centering}

		\begin{centering}
			\caption{\label{fig:Fig6}\textbf{Response Curves of a Single Oscillator}. We plot the response of a single oscillator in the coupled network (Fig.\ \ref{fig:Fig5}) in response to a range of forcing frequencies and 64 exponentially distributed input strengths (each colored curve corresponds to a different strength), $F=2^{-45}$ to $F=2^{18}$, which increase in the direction of the arrow from bottom to top. The two vertical dashed lines correspond to the true and effective characteristic frequencies.}
			\par\end{centering}
	\end{figure*}

	We consider the case of an $N=60$ oscillator network with characteristic frequencies $\omega_{i}=2^{\nicefrac{(i-1)}{10}}$ where $i\in \mathbb{Z}$,  $1\leqq i \leqq N$. We force every cell in the network with uniform input of equal strength and plot the response of a single oscillator as a function of forcing frequency $\omega$. We do this for a range of exponentially distributed forcings $2^{-45+k}$ where $k\in \mathbb{Z}, 0\leqq k \leqq 63$. The results are shown in Fig.\ \ref{fig:Fig6}, where each curve corresponds to a different forcing strength and the strength increase in the direction of the arrow from bottom to top.
	
	It is clear from Fig.\ \ref{fig:Fig6} that the oscillator's response is selective for a particular forcing frequency. For weak forcing, there is a single peak in the response curves, which corresponds to an \textit{effective characteristic frequency} different from the oscillator's characteristic frequency. This shift in frequency selectivity for weak forcing is a result of the oscillator integrating both direct input and input from other upstream oscillators with higher characteristic frequencies. For our specific model parameters, the effective characteristic frequency $\omega_{eff}$ is given by $2^{\frac{7}{10}}\omega_c$, where $\omega_c$ is the characteristic frequency of the oscillator. Note that $\omega_{eff}$ is exponentially distributed. As the forcing increases, the system reaches a regime in which the peaks of the response curves shift left towards the true characteristic frequency $\omega_{c}$, in agreement with previous studies \cite{Ruggero2000} and which can be seen in Fig.\ \ref{fig:Fig1}. Finally, for high forcing the curves flatten out across all frequencies, lose selectivity and compress by a factor of $1/3$. 
	
\begin{figure*}[bth]
	 	\begin{centering}
	 		\includegraphics[scale=0.6]{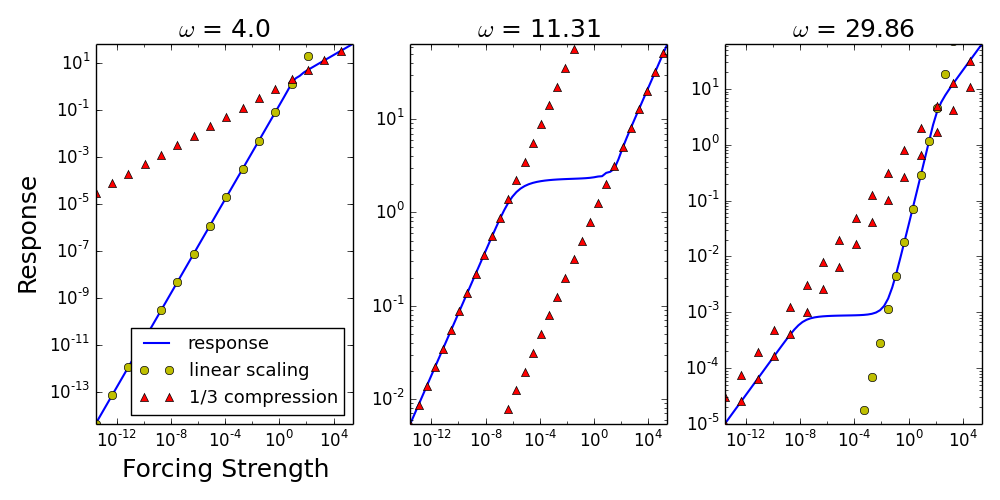}
	 		\par\end{centering}

		\begin{centering}
			
			\caption{\label{fig:Fig7}\textbf{Scaling Regimes in the Response of a Single Oscillator}. We plot the response of a single oscillator (Fig.\ \ref{fig:Fig6}) against the forcing strength for three different forcing frequencies $\omega$. This oscillator has characteristic frequency $\omega_{c}=8.00$ and effective frequency $\omega_{eff}=12.13$. These define the boundaries of three different scaling regimes. In the plots, the red triangles represent $1/3$ scaling, yellow dots represent linear scaling, and the blue curve is the oscillator response. In the first plot, $\omega=4.0$, a linear oscillator response turns into $1/3$ compression past a certain forcing strength. In the second regime, $\omega\approx11.3$, the response first scales as the cubic root of the input, then approaches a region of near invariance, whereafter it settles back down to a cubic root.  Finally, in the last regime, represented by $\omega\approx29.9$, we find the following scaling pattern: $1/3$, invariance, linear, $1/3$. }
			
			\par\end{centering}
		
	\end{figure*}

	We now examine the three frequency domains demarcated by the vertical dashed lines $\omega=\omega_{c}$ and $\omega=\omega_{eff}$ in Fig.\ \ref{fig:Fig6}. For $\omega<\omega_{c}$, the response scales linearly up until the point where the curves flatten out and compress by a factor of 1/3. For $\omega_{c}<\omega<\omega_{eff}$, we find a sharp transition to cubic root scaling for small forcing. This sharp transition is an artifact of the unidirectional coupling between oscillators; for forcing frequencies greater than the characteristic frequency, the oscillator integrates both direct input and input from oscillators upstream, while for forcing frequencies less than the characteristic frequency, the input from upstream oscillators is negligible compared to the direct input. For large forcing in this regime, ($\omega_{c}<\omega<\omega_{eff}$) we also find cubic compression. Between the two cubic scaling regimes, we observe the existence of higher-order compressive nonlinearities, with the response becoming nearly invariant as $\omega$ approaches $\omega_{eff}$. This invariance for medium forcing persists for the frequency domain $\omega>\omega_{eff}$ but is now accompanied by a large forcing regime in which the response scales linearly up until the curves flatten and exhibit cubic compression. The scaling behavior in the different frequency domains is depicted in Fig.\ \ref{fig:Fig7}. Our results generally agree with Fig.\ \ref{fig:Fig1} except for the sharp transition from linear to cubic scaling on the left flank of the response curves as the forcing frequency $\omega$ passes through $\omega_{c}$. 
	
	Finally, we note that weak input is preferentially amplified in the model. We define the amplification of an oscillator in our network as R/F  where $R$ is the oscillator response and $F$ the input amplitude. For $F<<1$ and forcing frequencies above $\omega_{c}$, $R\propto{F^{\nicefrac{1}{3}}}$, which implies that the amplification is $F^{-2/3}>>1$. To the left of $\omega_{c}$, there is no amplification of weak input.

	\subsection{Traveling Waves of Activity}
	
\begin{figure}[bth]
		\begin{centering}
			\includegraphics[scale=0.4]{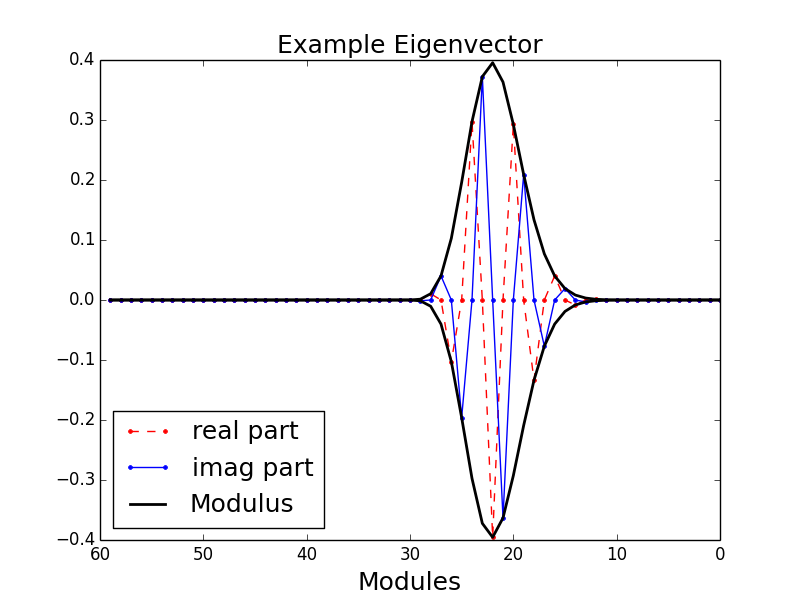}
			\par\end{centering}
		
		\begin{centering}
			
			\caption{\label{fig:Fig8}\textbf{An Example Eigenvector of the Network Connectivity Matrix A}. The real (red dashed) and imaginary (blue) parts of an eigenvector of the network connectivity matrix $A$ is plotted above along with the eigenvector's complex modulus. As waves in the model are determined by the linear part of (\ref{eq:hopfCompact}), wave behavior and shape can be studied in the context of eigenvectors.}
			
			\par\end{centering}
		
	\end{figure}

	In our toy model simulations, we observe traveling waves of activity, which propagate along the network of oscillators in the direction of decreasing characteristic frequency. Since the waves arise from the linear part of (\ref{eq:hopfCompact}), we can study the waves by examining the eigenvectors of the connectivity matrix $A$. In Fig.\ \ref{fig:Fig8}, we plot the real (red dashed) and imaginary (blue) parts of the eigenvector corresponding to the eigenfrequency of median absolute value. The envelope or complex modulus of the eigenvector is also depicted. We omit the final $N$ components of the eigenvector as they are a rescaled copy of the first $N$ components. 
	
	It is clear that the real and complex parts of the eigenvector are out of phase by $\pi/2$. This phase difference is the source of traveling wave behaviour in our model.  The envelope shape of the eigenvector is nearly symmetric and sharply peaks at a particular location, while the spatial frequency of the eigenvector is approximately uniform. These features are in disagreement with basilar membrane recordings, which predict a gradual increase in wave amplitude and spatial frequency in the direction of decreasing characteristic frequencies (basilar membrane base to apex).

	\subsection{Long-Range Connections}
\begin{figure}[bth]
		\begin{centering}
			\includegraphics[scale=0.28]{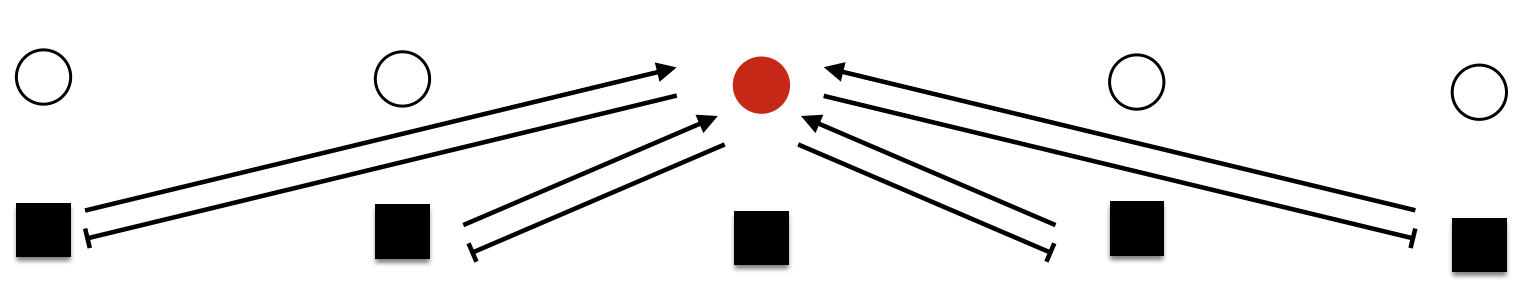}
			\par\end{centering}

		\begin{centering}
			
			\caption{\label{fig:Fig9}\textbf{Long-Range, Skew Symmetric Connectivity}. We illustrate the additional skew-symmetric connections to and from a single cell in the network. We omit showing the other connectivity (Fig.\ \ref{fig:Fig5}) to avoid clutter. Arrows denote excitation, while bars denote inhibition. This pattern is repeated for every cell in the network.}
			
			\par\end{centering}
		
	\end{figure}

	The addition of long-range connections between oscillators helps to resolve this disagreement by improving both wave shape and spatial frequency variation. We incorporate this into our toy model by including linear, long-range, skew-symmetric connections that are proportional in strength to the characteristic frequencies and decay as an exponential with distance. We define the strength of the connections between two oscillators separated a distance $d$ apart as $D{\omega}e^{-d}$ where $\omega$ is the characteristic frequency of the oscillator at the right side (higher frequency) of the connection and $D$ is a scaling parameter. The skew-symmetric connectivity is depicted in Fig.\ \ref{fig:Fig9}. These connections sit on top of the original connections from Fig.\ \ref{fig:Fig5}. 
	
\begin{figure*}[bth]
	 	\begin{centering}
	 		\includegraphics[scale=0.5]{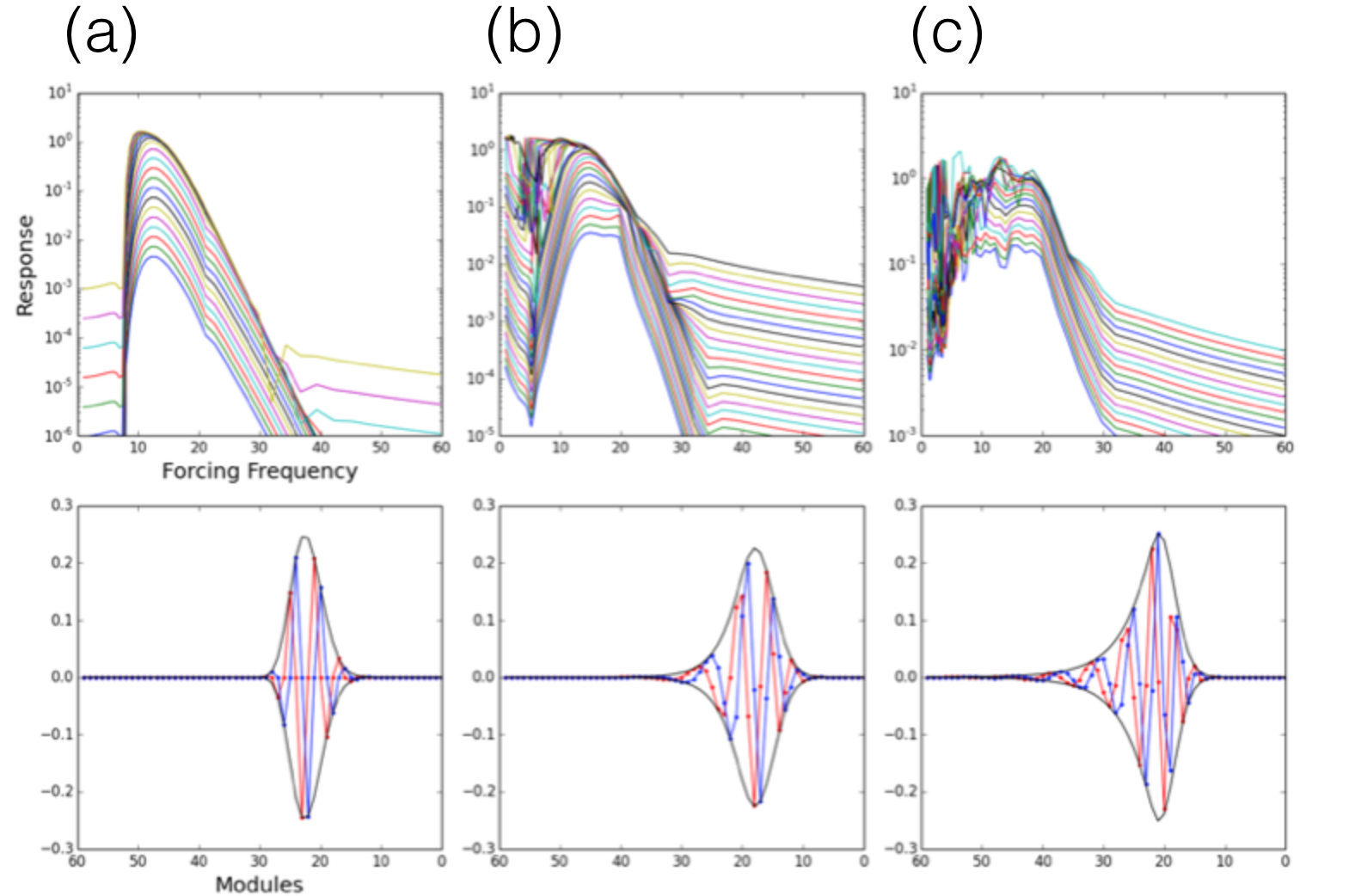}
	 		\par\end{centering}

		\begin{centering}
			
			\caption{\label{fig:Fig10}\textbf{Coupled Critical Oscillators with Long-Range Connections}. The response of a single oscillator in the network with long-range connections is plotted along with a representative eigenvector for three different scaling parameters $D$ of the skew-symmetric connections. Columns (a), (b), (c) correspond to $D=0$, $0.25$, $0.75$, respectively.}
			
			\par\end{centering}
		
	\end{figure*}

	We plot the response curves and eigenvectors of the system in Fig.\ \ref{fig:Fig10}(a) is just the orignal response curves without any skew-symmetric connections. Fig.\ \ref{fig:Fig10}(b), corresponding to $D=0.25$, exhibits a shift in the shape of the eigenvectors with the amplitude gradually increasing in the direction of decreasing characteristic frequencies.  The spatial frequency of the eigenvector also increases in the direction of decreasing characteristic frequency. The case of relatively strong skew-symmetric connections, $D=0.75$, is plotted in Fig.\ \ref{fig:Fig10}(c). The initial tail and spatial frequency variation are exaggerated in comparison to (a) and (b) and generally in agreement with basilar membrane studies, but this comes at a cost. As we increase $D$, the response curves lose their shape and no longer exhibit higher-order compression over a wide range of forcings; this is a direct result of the long range connections pushing the eigenvalues away from the imaginary axis and preventing the system from taking full advantage of the unique properties of center manifold dynamics. Therefore, to achieve response curves and waves that agree with experiments, a balance between the strength of the skew-symmetric and original network connections is needed.
	
	\section{Conclusion}
	
	We have shown that high-dimensional nonhyperbolicity with dynamics residing on a full-dimensional center manifold can exhibit the key characteristic nonlinearities of the active process in cochlea: frequency selectivity, amplification of weak input, and higher-order nonlinear compression. The toy model presented in this paper also gives rise to traveling waves of activity that propagate unidirectionally across the system. In order to better approximate the experimentally measured shape of traveling waves on the basilar membrane, we have included long-range, skew-symmetric connections between oscillators; however, these connections negatively impact the shape of the response curves by pushing the dynamics away from nonhyperbolicity. While we have been able to find a reasonable balance between the shape of traveling waves and response curves (Fig.\ \ref{fig:Fig10}(b)), we believe this conflict could be better solved by either finding an appropriate class of long range linear connections that preserve the high dimensionality of the center manifold or by including nonlinear couplings between the critical oscillators. Nonlinear couplings won't affect the linearization of the system, leaving the high-dimensional  nonhyperbolicity of the original system unchanged.
	
	As our model is just a toy model of center manifold dynamics, we do not suggest that our abstract critical oscillator network corresponds to actual anatomical and biophysical structures in the cochlea; although, if the cochlea does indeed utilize center manifolds in the processing of auditory stimuli, it might well be the case that the full high-dimensional phase space of cochlear dynamics could be reduced to a more simple structure on the center manifold. One possible simple structure is the toy model presented in this paper. 
	
	\section*{Acknowledgements} 
	We would like to thank Alex Katsov for useful discussions.  
	
	K.H. and D.M. contributed equally to this work.


\begin{thebibliography}{}
		
		\bibitem{Gold1948H1} T. Gold and R.J. Pumphrey, Proc. R. Soc. London B. \textbf{135}, 462 (1948).
		
		\bibitem{Gold1948H2} T. Gold, Proc. R. Soc. London B.  \textbf{135}, 492 (1948).
		
		\bibitem{Ruggero1992} M.A. Ruggero,  Curr. Opin. Neurobiol. \textbf{2}, 449 (1992).
		
		\bibitem{Ruggero2000} M.A. Ruggero et al, Proc. Natl. Acad. Sci. U.S.A. \textbf{22}, 11744 (2000).
		
		\bibitem{Bekesy1960} G. Von B\'{e}k\'{e}sy and E.G. Wever, Experiments in Hearing (McGraw-Hill
		Book Co., New York, 1960).
		
		\bibitem{Ruggero1997} M.A. Ruggero et al., J. Acoust. Soc. Am. \textbf{101}, 2151 (1997).
		
		\bibitem{Hudpseth2008} A.J. Hudspeth, Neuron \textbf{4}, 530 (2008).
		
		\bibitem{Hudspeth2000} A.J. Hudspeth et al., Proc. Natl. Acad. Sci. U.S.A. \textbf{97}, 11765 (2000).
		
		\bibitem{Martin2008} P. Martin, Active Processes and Otoacoustic Emissions in Hearing (Springer, 2008), pp. 93-143.
		
		\bibitem{Martin1999} P. Martin and A.J. Hudspeth, Proc. Natl. Acad. Sci. U.S.A. \textbf{96}, 14306 (1999).
		
		\bibitem{Martin2001} P. Martin and A.J. Hudspeth, Proc. Natl. Acad. Sci. U.S.A. \textbf{98}, 14386 (2001).
		
		\bibitem{Chan2005Dec} D.K. Chan and A.J. Hudspeth, Biophys. J. \textbf{89}, 4382 (2005).
		
		\bibitem{Chan2005Feb} D.K. Chan and A.J. Hudspeth, Nat. Neurosci. \textbf{8}, (2005).
		
		\bibitem{Kennedy2005} H.J. Kennedy et al., Nature \textbf{433}, (2005).
		
		\bibitem{Kennedy2006} H.J. Kennedy et al., J. Neurosci. \textbf{26}, 2757 (2006).
		
		\bibitem{Ashmore2008} J. Ashmore, Physiol. Rev. \textbf{88}, 173 (2008).
		
		\bibitem{Dallos2006} P. Dallos et al., J. Physiol. \textbf{576}, 37 (2006).
		
		\bibitem{Fettiplace2006} R. Fettiplace and C.M. Hackne, Nature Rev. Neurosci. \textbf{7}, 19 (2006).
		
		\bibitem{Hudspeth2010} A.J. Hudspeth et al., J. Neurophysiol.  \textbf{104}, 1219 (2010).
		
		\bibitem{Wiggins2003introduction} S. Wiggins, Introduction to Applied Nonlinear Dynamical Systems and Chaos (Springer, 1990).
		
		\bibitem{Kuznetsov2013} Y.A. Kuznetsov,  Elements of applied bifurcation theory. (Springer, 2013).
		
		\bibitem{Choe1998} Y. Choe, M.O. Magnasco, and A.J. Hudspeth, Proc. Natl. Acad. Sci. U.S.A. \textbf{95}, 15321 (1998).
		
		\bibitem{Eguiluz2000} V.M. Egu\'{i}luz et al.,   Phys. Rev. Lett.  \textbf{22}, 5232 (2000).
		
		\bibitem{Camalet2000} S. Camalet et al.,Proc. Natl. Acad. Sci. U.S.A.. \textbf{97}, 3183 (2000). 
		
		\bibitem{Magnasco2003} M.O. Magnasco, Phys. Rev. Lett. \textbf{90}, 058101 (2003).
		
		\bibitem{Duke2003} T. Duke and F. Jülicher, Phys. Rev. Lett. \textbf{90}, 158101 (2003).
		
		
		\bibitem{Kern2003} A. Kern and R. Stoop, Phys. Rev. Lett.  \textbf{91}, 128101 (2003).
		
		
		\bibitem{Grobman1959} D.M. Grobman, Dokl. Akad. Nauk SSSR. \textbf{128}, 880 (1959).
		
		\bibitem{Hartman1960a} P. Hartman, ‎Proc. Am. Math. Soc. \textbf{11}, 610 (1960).
		
		\bibitem{Hartman1960b} P. Hartman, Bol. Soc. Mat. Mex.  \textbf{5}, 220 (1960).
		
		\bibitem{Yan2012} X.H. Yan and M.O. Magnasco, PloS ONE.  \textbf{7}, e41419 (2012).
		
		\bibitem{Hayton2018} K. Hayton, D. Moirogiannis, and M. Magnasco, PLoS ONE \textbf{13}, e0196566 (2018).
		
		\bibitem{Carr2012} J. Carr, Applications of centre manifold theory (Springer, 2012).
		
		\bibitem{Kelley1967} A. Kelley, Journal of Differential Equations.  \textbf{3}, 546 (1967).
		
		\bibitem{ermentrout2010mathematical} G.B. Ermentrout and D.H. Terman, Mathematical foundations of neuroscience (Springer, 2010).
		
		\bibitem{hoppensteadt2012weakly} F.C. Hoppensteadt and E.M. Izhikevich,  Weakly connected neural networks (Springer, 2012).
		
		\bibitem{solovey2015} G. Solovey et al., J. Neurosci. \textbf{35}, 10866 (2015).
		
		\bibitem{churchland2012} M.M. Churchland et al., Nature.  \textbf{487}, 51 (2012).
		
		\bibitem{seung1998} H.S. Seung, Neural Networks.  \textbf{11}, 1253 (1998).
		
		\bibitem{seung2000} H.S. Seung et al., Neuron. \textbf{26},  259 (2000).
		
		\bibitem{machens2005} C.K. Machens, R. Romo, and C.D. Brody, Science \textbf{307}, 1121 (2005).
		
		\bibitem{freeman2005} W.J. Freeman and M.D. Holmes, Neural Networks. \textbf{18}, 497 (2005).
		
		\bibitem{bienenstock1998} E. Bienenstock and D. Lehmann,   Adv. Complex Syst. \textbf{1}, 361 (1998).
		
		\bibitem{beggs2012being} J.M. Beggs and N. Timme, Frontiers in physiology.   \textbf{3}, 163 (2012).
		
		\bibitem{chialvo2010emergent} D.R. Chialvo, Nat. Phys.  \textbf{6}, 744 (2010).
		
		\bibitem{mora2011biological} T. Mora and W. Bialek, J. Stat. Phys.  \textbf{144}, 268 (2011).
		
		\bibitem{da1998criticality} L. da Silva, A.R. Papa, and  A.C. de Souza, Phy. Rev. A. \textbf{242}, 343 (1998).
		
		\bibitem{fraiman2009ising} D. Fraiman et al., Phy. Rev. E.  \textbf{79}, 061922 (2009).
		
		\bibitem{beggs2003neuronal} J.M. Beggs and D. Plenz, J. Neurosci.  \textbf{23}, 11167 (2003).
		
		\bibitem{levina2007dynamical} A. Levina, J.M. Herrmann, and T. Geisel, Nat. Phys. \textbf{3}, 857 (2007).
		
		\bibitem{gireesh2008neuronal} E.D. Gireesh and D. Plenz, Proc. Natl. Acad. Sci. U.S.A. \textbf{105}, 7576 (2008).
		
		
		\bibitem{eguiluz2005scale} V.M. Eguiluz et al., Phys. Rev. Lett.  \textbf{94}, 018102 (2005).
		
		\bibitem{kitzbichler2009broadband} M.G. Kitzbichler et al., PLOS Comput. Biol. \textbf{5}, e1000314 (2009).
		
		\bibitem{magnasco2009self} M.O. Magnasco, O. Piro, and G.A. Cecchi, Phys. Rev. Lett. \textbf{102}, 258102 (2009).
		
		
		\bibitem{kanders2017avalanche} K. Kanders, T. Lorimer, and R. Stoop, Chaos \textbf{27}, 047408 (2017).
		
		
	\end{thebibliography}
\end{document}